\documentclass[prl, twocolumn, showpacs, superscriptaddress]{revtex4-2}
\usepackage{amsmath, amssymb}
\usepackage{graphicx}
\renewcommand{\section}[1]{{\par\it #1.---}\ignorespaces}

\begin{document}
\title{Characterizing Superradiant Phase of the Quantum Rabi Model}
\author{Yun-Tong Yang}
\affiliation{School of Physical Science and Technology, Lanzhou University, Lanzhou 730000, China}
\affiliation{Lanzhou Center for Theoretical Physics $\&$ Key Laboratory of Theoretical Physics of Gansu Province, Lanzhou University, Lanzhou 730000, China}
\author{Hong-Gang Luo}
\email{luohg@lzu.edu.cn}
\affiliation{School of Physical Science and Technology, Lanzhou University, Lanzhou 730000, China}
\affiliation{Lanzhou Center for Theoretical Physics $\&$ Key Laboratory of Theoretical Physics of Gansu Province, Lanzhou University, Lanzhou 730000, China}
\affiliation{Beijing Computational Science Research Center, Beijing 100084, China}


\begin{abstract}
Recently, a superradiant phase transition first predicted theoretically in the quantum Rabi model (QRM) has been verified experimentally. This further stimulates the interest in the study of the process of phase transition and the nature of the superradiant phase since the fundamental role of the QRM in describing the interaction of light and matter, and more importantly, the QRM contains rich physics deserving further exploration despite its simplicity. Here we propose a scheme consisting of two successive diagonalization to accurately obtain the ground-state and excited states wavefunctions of the QRM in full parameter regime ranging from weak to deep-strong couplings. Thus one is able to see how the phase transition happens and how the photons populate in Fock space of the superradiant phase. We characterize the photon populations by borrowing the distribution concept in random matrix theory and find that the photon population follows a Poissonian-like distribution once the phase transition happens and further exhibits the statistics of Gaussian unitary ensemble as increasing coupling strength. More interestingly, the photons in the excited states behave even like the statistics of Gaussian orthogonal ensemble. Our results not only deepen understanding on the superradiant phase transition but also provide an insight on the nature of the superradiant phase of the QRM and related models.
\end{abstract}
\maketitle

\section{Introduction}
Continually increasing couplings between different degrees of freedom in hybrid quantum systems provides a huge opportunity to explore new physics and/or new phenomena emerging from the interplay between the constituents of the hybrid systems \cite{Forn-Diaz2019, Kockum2019, Felicetti2020,Garcia-Vidal2021, Blais2021, Ashida2021}. In particular, after the strong coupling of a single photon to a superconducting qubit using circuit quantum electrodynamics has been firstly realized in 2004 \cite{Wallraff2004}, the ultrastrong coupling regime has been further achieved \cite{Niemczyk2010}, where the interaction energy is comparable to mode frequency, as a consequence, light and matter can mix together more tightly. Furthermore, the deep-strong couplings \cite{Casanova2010} has also been reached in which the interaction is even larger than the mode frequency \cite{Yoshihara2017, Yoshihara2018, Mueller2020}.

The achievement of strong hybridization not only leads to increased control of quantum systems \cite{Langford2017, Lv2018, Abah2020, Devi2020, Wang2020, Hastrup2021, Bin2021, Mei2022} and to possible applications on, e.g., lasers, quantum sensing \cite{Chu2021, Ilias2022}, and quantum information processing \cite{Romero2012, Lucas2018, Monroe2021, Head-Marsden2021}, but also provides chances to test many physical phenomena such as superradiance \cite{Hepp1973, Wang1973, Carmichael1973}, predicted theoretically in strong coupling regime of the Dicke model \cite{Dicke1954, Garraway2011}, even of the Rabi model \cite{Rabi1936, Rabi1937}. An early analysis of the ground state in the quantum Rabi model (QRM) showed that the ground state exhibits a squeezing in the deep-strong coupling regime \cite{Ashhab2010, Hwang2010}, a precursor of distinguished physics of superradiance, which has been further explored and confirmed theoretically \cite{De2013, Hwang2015, Ying2015, Liu2017, Wang2018Y, Peng2019, Zhu2020, Garbe2020, Jiang2021, Zhuang2021, Stransky2021}. Very recently this phase transition has been observed in a single trapped ion \cite{Cai2021} and stimulated experimentally in the platform of nuclear magnetic resonance \cite{Chen2021}.

Experimental observations of the superradiant phase transition further stimulate theoretical interest on the phase transition and the superradiant phase since the phases and phase transition are fundamental issues of modern condensed matter physics and related disciplines \cite{Chaikin1995}, in which the conventional paradigm is to identify the order parameter and broken symmetry associated under the framework of Landau phase transition theory \cite{Landau1980}. Due to the solvability of the QRM \cite{Braak2011, Chen2012}, it is possible to solve accurately for the wavefunction of the QRM in full parameter regime ranged from weak to strong, ultra-strong, even deep-strong ones, though it is not an easy thing \cite{Wolf2013}. Thus we can see how the superradiant phase transition happens and how the photons populate in the superradiant phase based on wavefunction since it contains all information of the system. Here we provide a scheme consisting of two successive diagonalization, where the first one is made exactly in the two-level space and the second is done in the truncated Fock space in a controllable way in a sense that the convergence depends on the size of the truncated Fock basis, which in the present situation even a PC is enough. We confirm this convergence by comparison with those obtained by numerical exact diagonalization (ED).

With accurate wavefunctions at hand, we study the process of phase transition and see how the photons populate in Fock space of the superradiant phase by changing the coupling strength from weak to strong ones. In normal phase it is found that only ground state is populated and thus there are no photons in the system, in agreement with what one expects. Around the phase transition point, the high level states begin to be excited, and some photons begin to populate on them. Here it is helpful to borrow the distribution concept in the random matrix theory\cite{Bohigas1984, Mehta2004}. This population is found to follow the Poissonian-like distribution. Further increasing coupling strength, the population behaves like the statistics of Gaussian unitary ensemble (GUE). At the same time an effective potential with a double-well forms gradually around the center of the harmonic potential. Physical reason for this transition from the Poissonian statistics-like to the GUE one-like is due to the formed effective potential barrier, which blocks the tunneling between the low-lying levels bounded at two displaced minima. According to this picture, the photons populated on higher energy levels beyond the barrier should exhibit different behavior. Indeed, one checks the population of photons in excited states, and finds that the photons follow exactly the statistics-like of Gaussian orthogonal ensemble(GOE). The sprectra obtained here not only deepen understanding of the superradiant phase of the QRM in strong coupling regimes \cite{Rossatto2017}, but also have a profound implication on the nature of the superradiant phase of the QRM and its variants. An interesting issue on the integrability of related models \cite{Braak2011, Belobrov1976, Milonni1983, Graham1984, Kuse1985, Bonci1991, Fukuo1998, Emary2003, Naether2014} deserves further investigation but is obviously beyond the scope of the present work.

\section{Model and Method}
The Hamiltonian of the QRM consists of a single photon mode and a two-level atom and their coupling, denoting by $H = H_0 + H_\sigma$, where $ H_0 = \hbar\omega a^\dagger a$ and $H_\sigma = \frac{\Delta}{2}\sigma_x + g\sigma_z (a + a^\dagger)$. Here $a^\dagger (a)$ is creation (destruction) operator of the single mode photon field and $\sigma_x, \sigma_z$ are usually Pauli matrices denoting the two-level atom. For convenience, we rescale the Hamiltonian by the mode frequency $\hbar\omega$, thus the two-level interval $\Delta$ and the coupling strength $g$ used in the following are dimensionless. It is also useful to use dimensionless position-momentum operators related to the destruction (creation) operator by $a = \frac{1}{\sqrt{2}}\left(\xi + \frac{\partial}{\partial \xi}\right)$  and  $a^\dagger = \frac{1}{\sqrt{2}}\left(\xi - \frac{\partial}{\partial \xi}\right)$ to rewrite the Hamiltonian as $
H_0 = \frac{1}{2} \left(-\frac{\partial^2}{\partial\xi^2} + \xi^2\right)$ and $H_\sigma = \frac{1}{2} \left(\Delta \sigma_x + 2\sqrt{2} g\sigma_z \xi\right)$ with a matrix form 
\begin{equation}\label{Hsigma_matrix}
H_\sigma =
 \frac{1}{2} \left(
  \begin{array}{cc}
    2\sqrt{2} g \xi & \Delta \\
    \Delta & -2\sqrt{2} g \xi \\
  \end{array}
\right).
\end{equation}
Thus Eq. \eqref{Hsigma_matrix} can be formally diagonalized and its eigenvales and eigenvectors read
\begin{eqnarray}
&& \epsilon_\pm(\xi) = \pm \frac{\Delta}{2} \sqrt{1 + \beta^2 \xi^2}, \label{eigenvalue}\\
&& \phi_{\pm}(\xi) = \frac{1}{\sqrt{2}} \left(\pm (1 \pm \gamma(\xi))^{\frac{1}{2}}, (1 \mp \gamma(\xi))^{\frac{1}{2}}\right)^T, \label{eigenvector}
\end{eqnarray}
where $\beta = \frac{2\sqrt{2} g}{\Delta}$ and $\gamma(\xi) = \frac{\beta \xi}{\sqrt{1 + \beta^2 \xi^2}}$. This finishes the first diagonization to solve the Schr\"odinger equation
\begin{equation}
H_\sigma \phi_\pm = \epsilon_\pm \phi_\pm, \label{sigma-diag}
\end{equation}
which includes the position as a parameter. However, our aim is to solve the full Hamiltonian $H$ satisfying with the full Schr\"odinger equation
\begin{equation}
H\Psi^E = (H_0 + H_\sigma)\Psi^E =  E\Psi^E
\end{equation}
To proceed, it is useful to assume two complete basis $|\xi,\sigma\rangle :=|\xi\rangle|\sigma\rangle$ and $|\xi,\pm\rangle := |\xi\rangle|\pm\rangle$, and one can write the total wavefunction $\Psi^E$ as $\Phi^E(\xi,\sigma) = \langle \xi,\sigma|\Psi^E\rangle$ or $\psi^E_\pm(\xi) = \langle\xi,\pm|\Psi^E\rangle$. Thus, one uses the orthogonal basis $ 1 = \sum_\pm \int d\xi |\xi,\pm\rangle\langle\xi,\pm|$,
\begin{eqnarray}
\Phi^E(\xi,\sigma) &=& \sum_\pm \int d\xi' \langle \xi,\sigma|\xi',\pm\rangle\langle\xi',\pm|\Psi^E\rangle  \nonumber\\
&=& \sum_\pm \langle \sigma|\pm\rangle \langle\xi,\pm|\Psi^E\rangle \nonumber \\
&=& \sum_\pm \phi_\pm(\xi) \psi^E_\pm(\xi) \label{schrodinger2}
\end{eqnarray}

To consider the Born-Oppenheimer approximation \cite{Born1927}, one has
\begin{equation}
(H_0 + H_\sigma) \phi_\pm(\xi) \psi^E_\pm(\xi) = E \phi_\pm(\xi) \psi^E_\pm(\xi) \label{Schrodinger3}
\end{equation}
Multiplying by $\phi_\pm^*$ Eq. (\ref{Schrodinger3}), one obtains
\begin{equation}
(H_{0,\pm} + \epsilon_\pm)\psi^E_\pm(\xi) = E_\pm \psi^E_\pm(\xi), \label{Schrodinger4}
\end{equation}
where $H_{0,\pm} = \phi^*_\pm H_0 \phi_\pm = H_0$ which is easy to verify. Eq. (\ref{Schrodinger4}) is the main result of the present work, which is the starting point of the following calculation.

In order to solve Eq. (\ref{Schrodinger4}), one inserts the complete basis $1 = \sum_n|n\rangle\langle n|$ of the standard harmonic oscillator into Eq. (\ref{Schrodinger4}) to obtain
\begin{eqnarray}
\sum_{m}\langle n|(H_0 + \epsilon_\pm)|m\rangle\langle m|\psi^E_\pm\rangle = E_{\pm,n} \langle n|\psi^E_\pm\rangle. \label{Schrodinger5}
\end{eqnarray}
In a truncated basis $|n\rangle, n = 0, 1, \cdots, N-1$, solving Eq. (\ref{Schrodinger5}) is equivalent to diagonalize the following $N\times N$ matrix
\begin{equation}
\left(
  \begin{array}{cccc}
    \langle 0|H_0+\epsilon_\pm|0\rangle & \cdots & \langle 0|\epsilon_\pm|N-1\rangle \\
    \vdots & \ddots & \vdots \\
    \langle N-1|\epsilon_\pm|0\rangle & \cdots & \langle N-1|H_0+\epsilon_\pm|N-1\rangle \\
  \end{array}
\right)
\end{equation}
This finishes the second diagonalization, which gives the spectra of the corresponding wavefunction. The convergence and its accuracy are verified in comparison with the numerical ED, as given in SM \cite{sm2022}. In the following we focus on the ground state and low-lying excited states given by the negative branch and the positive one has much high energy under the parameters we use below and will be explored in future. 

\section{Results and Discussion}
\begin{figure}[h]
\begin{center}
\includegraphics[width = \columnwidth]{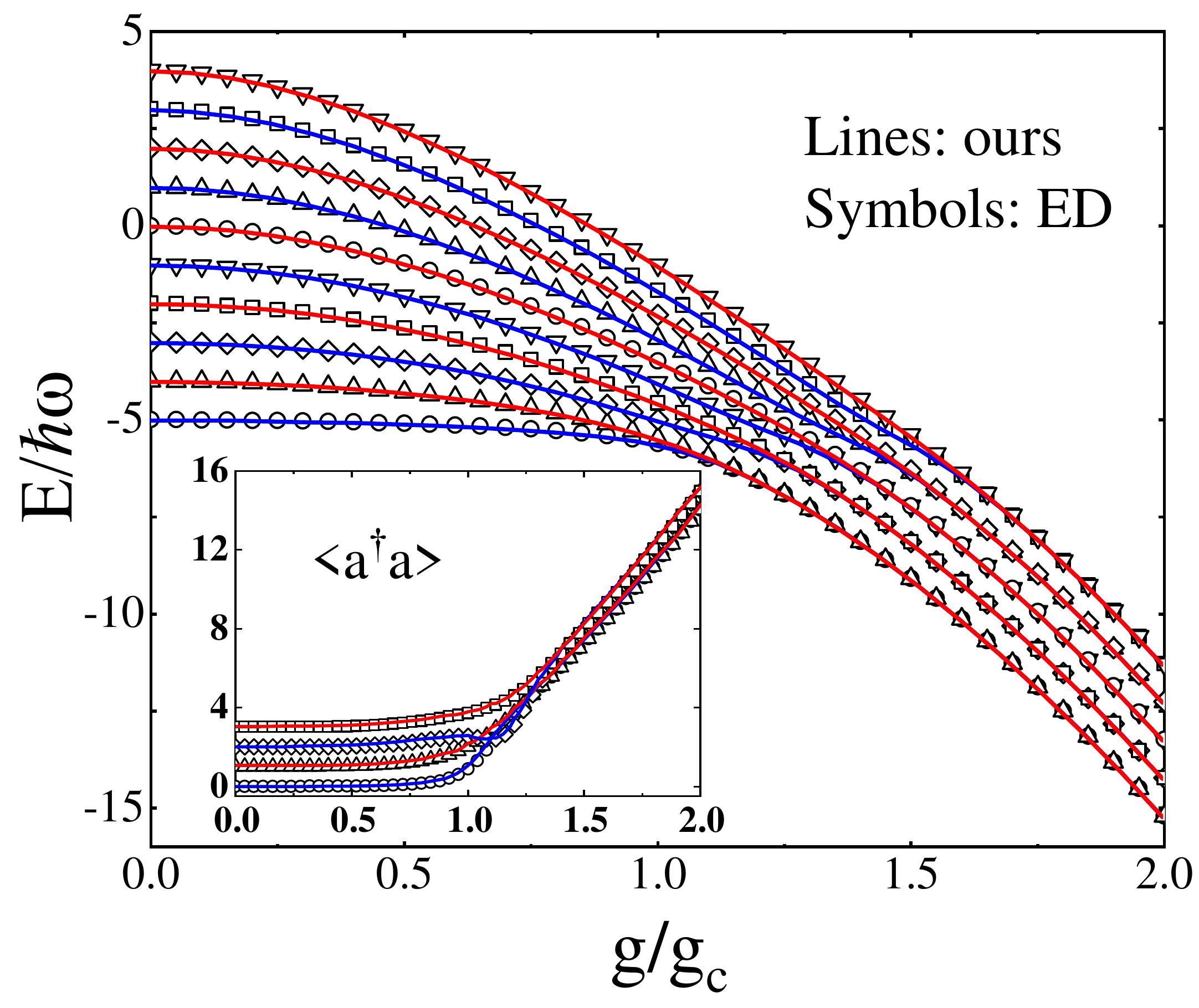}
\caption{The energy levels of the ground state and the other $9$ low-lying excited states as functions of the coupling strength scaled by $g_c = \sqrt{1+\sqrt{1+\frac{\Delta^2}{16}}}$ \cite{Ying2015}. The lines (blue and red ones denote different parity) are our results and the symbols those obtained by numerical ED with the same parameter $\Delta = 10$. The inset presents the result of photons as functions of the coupling strength for the ground-state and the first three excited states. }\label{fig1}
\end{center}
\end{figure}

Figure \ref{fig1} shows the ground state and the low-lying excited states energies (blue and red lines with odd and even parity, respentively) as functions of coupling strength scaled by $g_c = \sqrt{1+\sqrt{1 + \frac{\Delta^2}{16}}}$ roughly marked the superradiant phase transition point for the ground state \cite{Ying2015}. To verify the accuracy of the present calculation, we also provide the results marked by symbols obtained by numerical ED for the same model parameter $\Delta = 10$. One sees that ranged from weak to strong coupling regimes, our results are in excellent agreement with the exact ones in a high precision. The phase transitions are found to happen smoothly from normal phase to superradiant phase. This is in contrast to some previous variational methods \cite{Yu2012, Liu2015, Ying2015, Cong2017, Mao2019, Sun2020, Li2021b} or approximations \cite{Irish2007,Chen2020,Li2021}. One can also refer to recent overviews on the related issues \cite{Braak2019, LeBoite2020}. The superradiant phase transitions also happen in the low-lying excited states, but with larger coupling strengths, which are consistent with those reported in literature \cite{Puebla2016}. The inset shown in Fig. \ref{fig1} gives the photons as functions of the coupling strength, which go up at the point of phase transition and further increase with increasing coupling strength.

\begin{figure}[h]
\begin{center}
\includegraphics[width = \columnwidth]{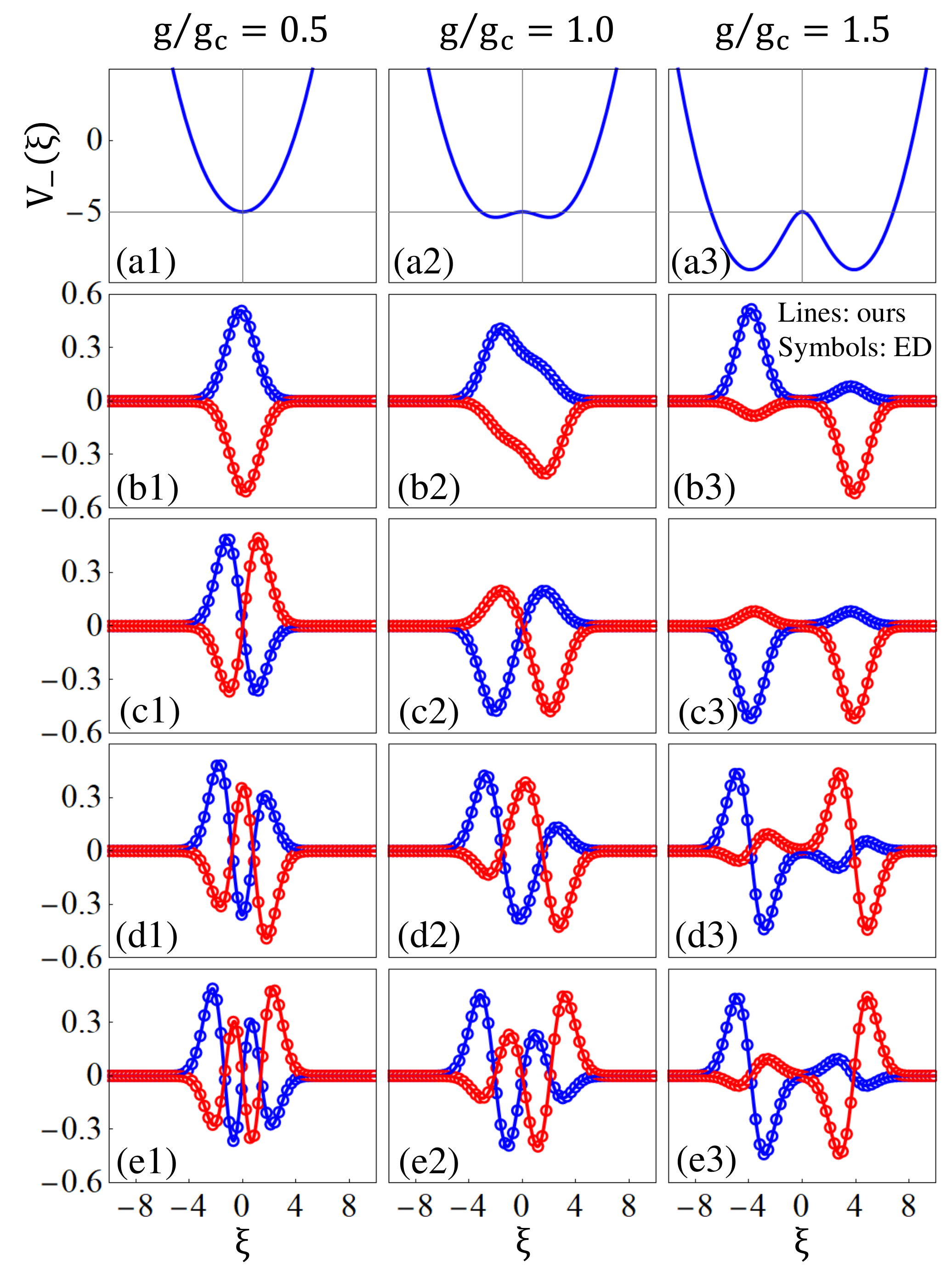}
\caption{The effective potentials (a1-a3) and the wavefunctions (blue and red lines) for the ground state (b1-b3) and the first (c1-c3), the second (d1-d3), and the third (e1-e3) excited states for three coupling strength $g/g_c = 0.5$, $1.0$, and $1.5$. For comparison, we also present the results of numerical ED (symbols). The parameters used are the same as in Fig. \ref{fig1}. From the wavefunctions obtained the parity of the states is not obviously broken. }\label{fig2}
\end{center}
\end{figure}

Impressive accuracy of our method can be further confirmed by wavefunctions presented in Fig. \ref{fig2}, in which the ground state wavefunction and those of the first three low-lying excited states are plotted for three coupling strengths $g/g_c = 0.5, 1.0, $ and $1.5$, which correspond to the normal phase, roughly superradiant phase transition point, and the superradiant phase, respectively. Likewise, we also present the results obtained by numerical ED. The lines (red and blue ones denote the two components of the wavefunctions) denote our results and the symbols those of ED. Fig. \ref{fig2}(a1-a3) show the effective potential given exactly by $V_{-}(\xi) = \frac{1}{2}\xi^2 - \frac{\Delta}{2} \sqrt{1 + \beta^2 \xi^2}$, from which in weak-coupling regime, the quantity $\beta$ is small, thus the potential is roughly the standard harmonic oscillator potential, and the local minimum locates at the point of $\xi = 0$. Increasing the coupling strength up to the critical point, the local minimum begins to become local maximum, and a tiny ``Mexician Cap" forms, accompanying with a separation of the ground state wavefunction, as shown in Fig. \ref{fig2}(b2). A complete separation of the wave-packets is observed in Fig. \ref{fig2} (b3), which represents that the system enters completely into the superradiant phase. Correspondingly, an effective double-well potential \cite{Irish2014} develops well, as shown in Fig. \ref{fig2}(a3). This double-well potential can also be obtained by a Taylor expansion
\begin{equation}
V_{-}(\xi) \approx -\frac{\Delta}{2} + \left(\frac{1}{2} - \frac{\beta^2\Delta}{4}\right)\xi^2 + \frac{\beta^4\Delta}{16}\xi^4, \label{v-expansion}
\end{equation}
which corresponds to the standard form in the Landau phase transition theory except for a constant energy.  This double-well potential plays an important role in the photons population, as discussed later. Another important feature can be observed from the wavefunctions including the ground state and the excited states, namely, an obviously breaking of parity is absent up to the coupling strength $g/g_c = 1.5$. We believe that this result is correct since it is also absent in numerical ED. Physical reason is still unclear and we would like to leave for future study. 
\begin{figure}[h]
\begin{center}
\includegraphics[width = \columnwidth]{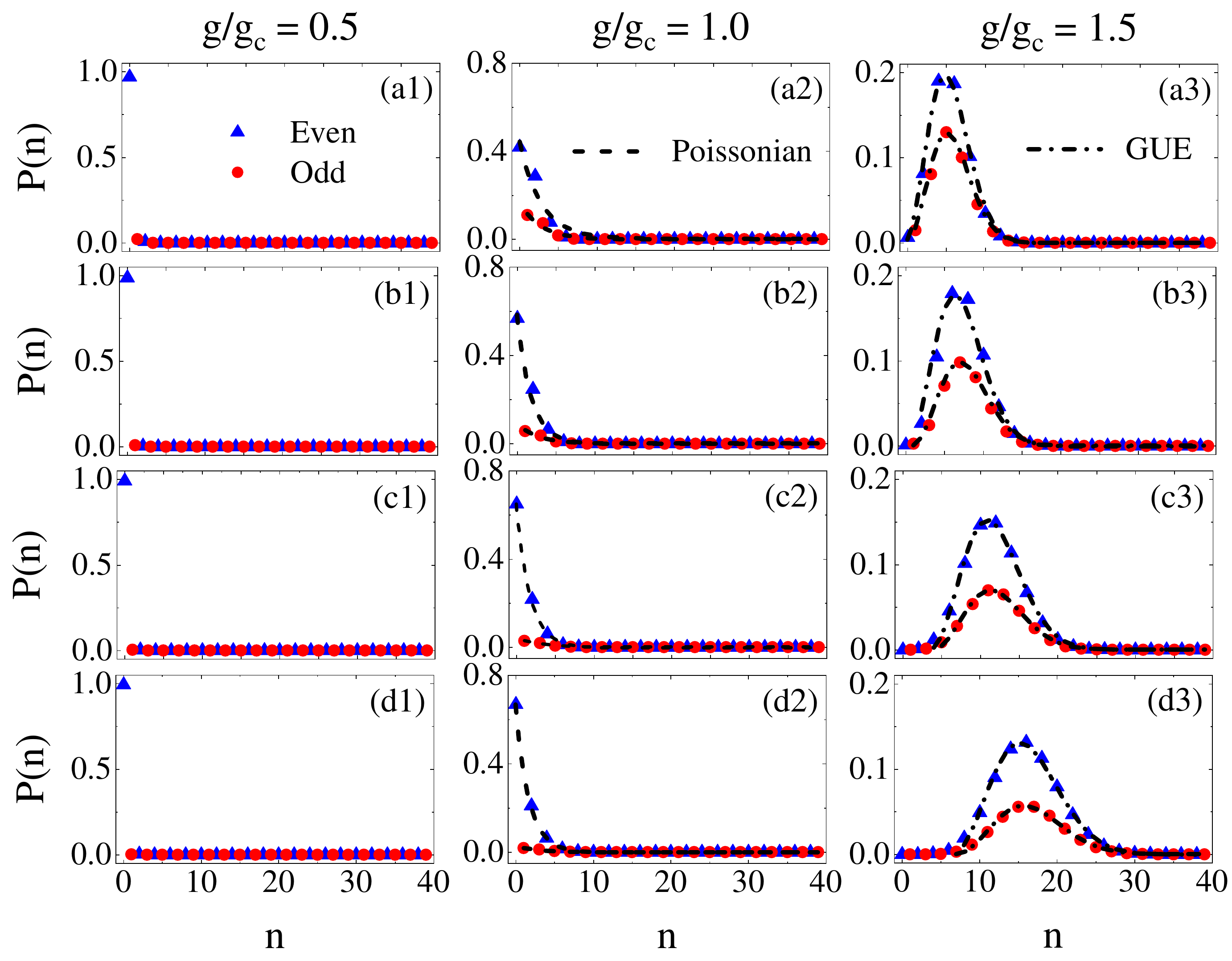}
\caption{The photon population $P(n)$ in Fock space for the ground state in three coupling strengths $g/g_c = 0.5, 1.0$, and $1.5$ for different $\Delta$'s: $5$ (the first row, a1-a3), $10$ (the second row, b1-b3), $20$ (the third row, c1-c3), and $30$ (the fourth row, d1-d3). The red dots and blue triangles denote the photon population in Fock basis with odd and even parity. The dash lines represent the fits of the Poissonian statistics-like and the dashed-dot lines represent the fits of the statistics of GUE-like. The fitting parameters and details are given in SM \cite{sm2022}.}\label{fig3}
\end{center}
\end{figure}

Next we move to the photon population calculated only from the wavefunctions. Fig. \ref{fig3} shows the results for three typical regimes: the normal phase ($g/g_c = 0.5$), the superradiant phase transition point ($g/g_c = 1.0$), and the superradiant phase ($g/g_c = 1.5$) and for four different $\Delta$'s: $5$ ( the first row, a1-a3), $10$ (the second row, b1-b3), $20$ (the third row, c1-c3), and $30$ (the fourth row, d1-d3). The first column is in the normal phase, in which no photons are excited except for a tiny upward observed in Fig. \ref{fig3}(a1) with small $\Delta = 5$. The second column denotes the situation around the superradiant phase transition point, in which the high levels begin to be excited. An obvious difference of photon numbers in odd and even channels of Fock space is observed. The even parity channels are more easier to excite than those with odd parity. The reason is that the ground state of the system (single mode photon plus the two-level) is odd parity, therefore the Fock basis with even parity meet this requirement. The third column represents the situation of the superradiant phase, in which Fock basis with more higher energy levels are excited. This is physically reasonable since with increasing coupling strength the photon number also increases, as also observed in experiment \cite{Cai2021}. It is noticed that in the superradiant phase the population of the low-lying energy levels are suppressed. As also mentioned above, this is due to the fact that an effective potential barrier forms with increasing the coupling strength, as shown in Fig. \ref{fig2}(a3). It blocks the tunneling of the states of low-lying energy levels around two separated local potentials. 

In order to further characterize the photon populations, it is helpful to borrow the distribution concept in random matrix theory \cite{Bohigas1984, Mehta2004}, from which it is well-known that there are three typical distributions
\begin{eqnarray}
&& P_P(s) = e^{-s}, \label{Poisson} \\
&& P_{GUE}(s) = \frac{32}{\pi^2} s^2 e^{-\frac{4s^2}{\pi}}, \label{gue} \\
&& P_{GOE}(s) = \frac{\pi}{2}s e^{-\frac{\pi s^2}{4}}, \label{goe}
\end{eqnarray}
which correspond to the Poissonian statistics, the statistics of GUE and that of GOE \cite{Amann2016}. In its original definition, the variable $s$ denotes the energy intervals of adjacent levels \cite{Berry1977}, but here we replace it by the Fock basis. We fit the photon population by the above formulas and the details are presented in SM \cite{sm2022}. The results are plotted in Fig. \ref{fig3} for the Poissonian-like statistics (dash lines) and the statistics of GUE-like (dashed-dot lines). The fitting is found well for all photon populations. Around the superradiant phase transition point, the both populations for odd and even components follow the Poissonian-like statistics. In the superradiant phase the photon populations become the statistics of GUE-like. The populations in low-lying energy levels are strongly suppressed due to the emerged potential barrier, as pointed out above.

\begin{figure}[h]
\begin{center}
\includegraphics[width = \columnwidth]{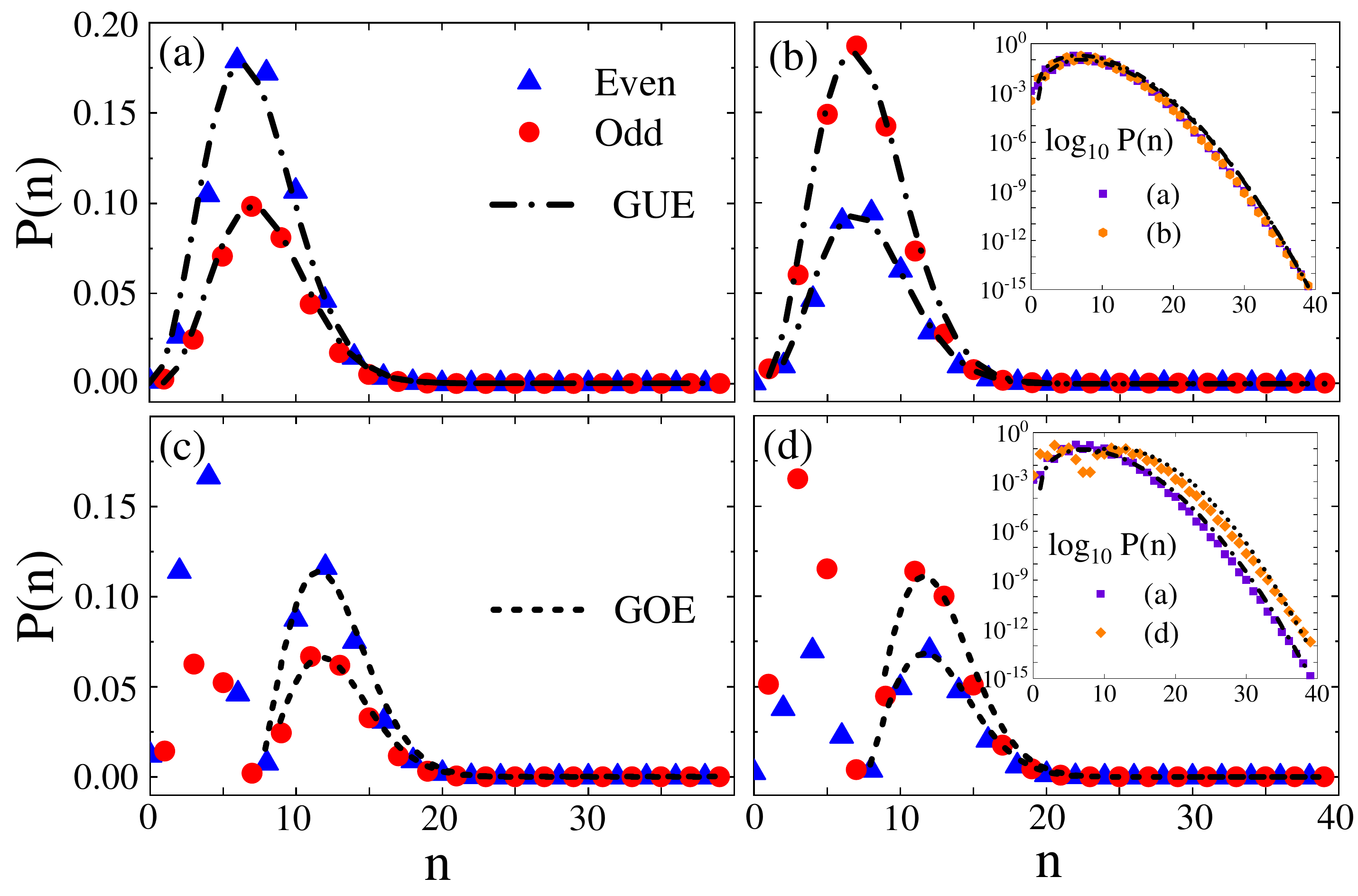}
\caption{The photon population of (a) the ground state, (b) the first, (c) the second, and (d) the third excited states in the superradiant phase. The parameters used read $\Delta = 10$ and $g/g_c = 1.5$. The insets in (b) and (d) show the details of the tails of GUE and GOE, in which the GUE decays more rapidly than that of the GOE. In the second and third excited states the populations of lower low-lying energy levels are scattered and do not exhibit typical statistical features.}\label{fig4}
\end{center}
\end{figure}
The same picture can also be applied to understand the photon population of the superradiant phase in the excited states of the system, as shown in Fig. \ref{fig4}, in which Fig. \ref{fig4}(a) is the same as that in Fig. \ref{fig3}(b3) and replotted here for comparison. The first excited state has a behavior of GUE-like, which indicates the first excited state is still influenced by the emerged potential barrier. More interesting is that the second and the third excited states behave like GOE. The inset in Fig. \ref{fig4}(d) shows the details of the tails in comparison to that of GUE. This transition of the populations is obviously due to the emerged double-well potential, by which the third and the fourth excited states should have a higher energy level than the induced potential barrier, therefore, its decay is more slower than those in the ground state and the first excited state. 

\section{Summary and Outlook}
We propose a scheme consisting of two successive diagonalization to obtain accurately wavefunctions of both ground state and excited states of the QRM in full parameter space ranged from weak to strong, ultra-strong, even deep-strong regimes in a controllable way. Based on the wavefunctions obtained, we characterize the process of superradiant phase transition and the nature of the superradiant phase. In particular, the photons population can be well characterized by the distributions borrowed from the random matrix theory, namely, Poissonian statistics, the statistics of GUE and that of GOE, dependent of the coupling strength and the excited states. 

In the present work we focus on the wavefunctions of the ground state and the low-lying excited states and characterize their photon populations. Although we borrow the distribution concept in random matrix theory, we do not touch the issue of integrability of the QRM \cite{Braak2011, Braak2019, Milonni1983, Graham1984, Kuse1985, Bonci1991, Fukuo1998} which involves the entire spectrum of the QRM. However, it keeps interest to explore the implication of the different photon populations and their transitions on the integrability of the QRM since the standard analysis of level statistics of the QRM is not sufficient to judge whether or not the QRM is integrable \cite{Kuse1985}.

\section{Acknowledgments}
The authors acknowledge Miaomiao Zhao and Junpeng Liu for the availability of numerical ED code and Liang Huang, Qing-Hu Chen, Hang Zheng and D. Braak for helpful discussions and valuable suggestions. The work is partly supported by the programs for NSFC of China (Grant No. 11834005, Grant No. 12047501).

%\bibliography{myrefs}

%\begin{thebibliography}{99}
%%\bibitem{Greenstein2006} G. Greenstein and A. Zajonc, \textit{The Quantum Challenge: Modern Research on the Founddations of Quantum Mechanics} (Jones and Bartlett Publishers, Sudbury, Massachusetts, 2006).
%%\bibitem{Laloe2019} F. Lalo\"e, \textit{Do We Really Understand Quantum Mechanics?} 2nd Ed. (Cambridge Univ. Press, 2019).
%%\bibitem{Pusey2012} M. F. Pusey, J. Barrett, and T. Rudolph, ``On the reality of the quantum state", Nat. Phys. \textbf{8}, 475-478 (2012).
%%\bibitem{Feynman1989} R. Feynman, \textit{The Feynman Lectures on Physics} (Addison-Wesley, Boston, 1989).
%%\bibitem{Dirac1958} P. A. M. Dirac, \textit{The Principle of Quantum Mechanics} (Oxford University Press, 1958), P74.
%apsrev4-2.bst 2019-01-14 (MD) hand-edited version of apsrev4-1.bst
%Control: key (0)
%Control: author (8) initials jnrlst
%Control: editor formatted (1) identically to author
%Control: production of article title (0) allowed
%Control: page (0) single
%Control: year (1) truncated
%Control: production of eprint (0) enabled
%

%\end{thebibliography}

\end{document}